\documentclass{nature}

\usepackage{graphicx}
\makeatletter
\let\saved@includegraphics\includegraphics
\AtBeginDocument{\let\includegraphics\saved@includegraphics}

\makeatother

\usepackage{multirow}
\usepackage{mathtools}
\usepackage[colorlinks=true,linkcolor=blue,citecolor=black,urlcolor=blue]{hyperref}
\usepackage[dvipsnames]{xcolor}
\usepackage{soul}
\usepackage{authblk}
\usepackage{lineno}
\usepackage[labelfont=bf]{caption}
\usepackage[figurename=Figure]{caption}
\usepackage{siunitx}
\usepackage{booktabs}
\usepackage{subfig}
\usepackage[T1]{fontenc}

\newcommand{\MUB}{\mu_{\text{B}}}

\newcommand{\mdsd}[1]{\textcolor{YellowOrange}{\textbf{[MdSD: #1]}}}
\newcommand{\filipe}[1]{\textcolor{red}{\textbf{[FG: #1]}}}

\newcommand{\new}[1]{\textcolor{black}{#1}}

\begin{document}

\title{Ultrafast light-induced magnetization in non-magnetic films: from orbital and spin Hall phenomena to the inverse Faraday effect}

\author[1,2]{Hanan Hamamera}

\author[3]{Filipe Souza Mendes Guimar\~aes}

\author[1,4,5]{Manuel dos Santos Dias}

\author[1,4,*]{Samir Lounis}

\affil[1]{Peter Gr\"unberg Institut and Institute for Advanced Simulation, Forschungszentrum J\"ulich \& JARA, 52425 J\"ulich, Germany}
\affil[2]{Department of Physics, RWTH Aachen University, 52056 Aachen, Germany}
\affil[3]{J\"ulich Supercomputing Centre, Forschungszentrum J\"ulich \& JARA, 52425 J\"ulich, Germany}
\affil[4]{Faculty of Physics, University of Duisburg-Essen \& CENIDE, 47053 Duisburg, Germany }
\affil[5]{Scientific Computing Department, STFC Daresbury Laboratory, Warrington WA4 4AD, United Kingdom}
\affil[*]{s.lounis@fz-juelich.de}

\maketitle


\begin{abstract}
The field of orbitronics has emerged with great potential to impact information technology by enabling environmentally friendly electronic devices. The main electronic degree of freedom at play is the orbital angular momentum, which can give rise to a myriad of phenomena such as the orbital Hall effect (OHE), torques and orbital magnetoelectric effects. Here, we explore via realistic time-dependent electronic structure simulations the magnetic response of a non-magnetic material, an ultrathin Pt film, to ultrafast laser pulses of different polarizatons and helicities. We demonstrate the generation of significant orbital and spin magnetizations and identify the underlying mechanisms consisting of the interplay of the OHE, inverse Faraday effect and spin-orbit interaction.  Our discoveries advocate for the prospect of encoding magnetic information using light in materials that are not inherently magnetic.
\end{abstract}

\maketitle 

\section*{Introduction}

Orbitronics~\cite{bernevig2005orbitronics,Kontani2007,Kontani2009,go2018intrinsic} pioneers a new research area, laying the foundation for advancements in device technology~\cite{Rappoport2023}. It utilizes orbital currents as a means of information transfer, offering a novel path for spintronics and valleytronics~\cite{Cysne2021}, as unveiled in recent experiments~\cite{ElHamdi2023,Choi2023,Hayashi2023,Seifert2023}. The origin of orbital currents can be traced to the orbital motion of electrons, generating orbital angular momentum~\cite{Go2021}. The control of this momentum in solids holds potential for information processing~\cite{Rappoport2023}. The crucial role of orbital currents extends to deciphering the complex traits of topological materials~\cite{Costa2023}, potentially impacting diverse orders~\cite{Cysne2023}, and engaging with quasi-particle excitations in unique ways~\cite{Lee2021}. This exploration holds the promise of solving puzzles in correlated matters and discovering remarkable quantum phenomena.

One of the key phenomena in orbitronics is the orbital Hall effect (OHE), which consists on the flow of the orbital angular momentum defined around atoms at each lattice to a perpendicular direction of an external electric field~\cite{Kontani2007,go2018intrinsic}. It is then proposed that it is with spin-orbit coupling \new{(SOC)}, which couples the spin and orbital degrees of freedom of the electrons, that the spin Hall effect (SHE)~\cite{Sinova2015} emerges out of the OHE~\cite{go2018intrinsic}. Upon application of an electric field $\mathbf{E}$, a charge current density $\mathbf{J_C} = \boldsymbol{\sigma}_0 \mathbf{E}$ is generated, where $\boldsymbol{\sigma}_0$ is the conductivity tensor of the system. 
The flow of the charge current through the heavy metal triggers via the \new{SOC}  a spin current $\mathbf{J}_s = \theta_\text{SH} \mathbf{J}_C \times \boldsymbol{\sigma}$ with polarization $ \boldsymbol{\sigma}$, where $\theta_{\text{SH}}$ represents the spin Hall angle which measures the charge-to-spin conversion rate. Along the out-of-plane direction $\hat{z}$, the flowing spin angular momentum  can accumulate if there is any edge, which induces a non-equilibrium spin polarization $\delta \mathbf{M}_{s} \propto \theta_\text{SH} \boldsymbol{\sigma}_0  \boldsymbol{\hat{z}}\times \mathbf{E}$. 
It is the current induced spin polarization that gives rise to the so-called spin-orbit torques (SOTs)~\cite{Garello2013}. Similar to the spin polarization, the orbital polarization can be related to the electric field $\delta \mathbf{M}_{L} \propto   \hat{\mathbf{L}}\times \mathbf{E}$ as a result of the orbital polarization current $\mathbf{J}_L \propto \mathbf{J}_C \times \hat{\mathbf{L}}$~\cite{Kontani2007,Kontani2009}. \new{The OHE can reach gigantic amplitudes as predicted theoretically~\cite{Jo2018}. As shown in a thorough experimental investigation~\cite{Sala2022}, the intertwining of the SHE and OHE can be rather complex in different materials and interfaces, which leads to rich physics. } 
While the OHE is a dynamical fingerprint of the motion of electrons, it is appealing to explore the potential of controlling the underlying effects at ultrafast timescales.

The realm of controlling magnetic states of matter at ultrafast timescales has been thoroughly explored since the demonstration of the discovery of optically driven ultrafast demagnetization~\cite{beaurepaire1996ultrafast}. The ultimate control of the magnetic behavior of materials at femtosecond (or even faster) timescales defines one of the most pursued paradigm shifts for future information technology~\cite{stanciu2007all,lambert2014all,John2017}. \new{Successful demonstrations addressing orbitronics at ultrafast time scales are scarce both experimentally~\cite{Seifert2023} and theoretically~\cite{Busch2023}.}

In our work, we demonstrate the all-optical generation of orbital and spin magnetizations in an initially non-magnetic material consisting of a Pt thin film utilizing single laser pulses. Pt is a transition metal element characterized by a large Stoner susceptibility~\cite{Nieuwenhuys1975,Khajetoorians2013,Meier2011,Bouhassoune2014,Bouhassoune2016,Azpiroz2017,Hermenau2017,bouaziz2020}, which means that it can easily carry a spin moment in the vicinity of magnetic atoms. It was also demonstrated that it can develop a large spin-polarization cloud around adatoms, which impact on both the magnitude and sign of their magnetic anisotropy energy~\cite{Bouhassoune2016} that has been probed indirectly via state-of-the-art \new{scanning tunneling microscopy (STM)} measurements for an Fe adatom on Pt(111) surface~\cite{Khajetoorians2013}. In the dynamical regime, the Stoner criterion can be fulfilled for some transition metal adatoms~\cite{Azpiroz2017}. In the current investigation, we utilize a time-dependent tight-binding framework parameterized from density functional theory (DFT) calculations (see Method section and Ref.~\citenum{Hamamera2022}), capable to explore non-linear magnetization dynamics up to a few picoseconds. 

\new{The method is implemented in a state-of-the-art code TITAN, which is developed to address various phenomena based on a realistic description of the electronic structure. TITAN stands for TIme-dependent Transport and Angular momentum in Nanostructures~\cite{titan_zenodo,Guimaraes2017,Guimaraes2019,Guimaraes2020}. Besides real-time ultrafast magnetization dynamics induced by the interaction with laser pulses~\cite{Hamamera2022}, TITAN can be used to explore  dynamical magnetic linear responses\cite{Guimaraes2015}, dynamical transport and torques\cite{Guimaraes2017,Guimaraes2020}, magnetic damping~\cite{Guimaraes2019} as well as superconducting-magnetic interfaces by solving the Bogoliubov--de Gennes equations self-consistently~\cite{Aceves2023,Aceves2024}.}

\new{We illustrate that substantial spin and orbital magnetizations can arise in Pt through strategic laser pulse design. The spin moments attain a magnitude comparable to that observed when Pt is directly interfaced with a magnetic layer. Remarkably, the orbital moments exhibit a similar scale to the spin moments, see e.g.~\cite{Khajetoorians2013,Bouhassoune2016}.}  
We identify an emergent anisotropy of the magnetization that depends on the polarization of the incident light pulses, driven by the interplay of the orbital Hall and spin Hall effects intertwined with the inverse Faraday effect. 
The inverse Faraday effect (IFE) was proposed as a direct mechanism for laser-induced demagnetization~\cite{kimel2005ultrafast}, light-induced magnetic torques~\cite{tesavrova2013experimental,zhang2016switching,Berrita2016,freimuth2016laser,Choi2017b,Hamamera2022} with an induced magnetization expected to be proportional to $\mathbf{E}\times\mathbf{E^*}$, where $\mathbf{E}$ is the complex electric field defined by the laser polarisation. Here, we dissect the distinct underlying mechanisms for the induced magnetizations and propose an experimental protocol to verify our predictions, which promote the potential of light-encoding of magnetic information in non-magnetic materials.

\section{Results}
We conduct tight-binding simulations with parameters derived from DFT. This involves the real-time propagation of ground state eigenvectors, achieved by solving the time-dependent Schrödinger equation for durations up to a few hundreds of femtoseconds~\cite{Hamamera2022}. The Hamiltonian on which the calculations are based encompasses the electronic hopping-derived kinetic energy, the SOC, interactions arising from electron–electron exchange, and the impact of excitations induced by laser irradiation.  The propagated solutions were then used to calculate the orbital and spin magnetizations after applying both linearly and circularly polarized laser pulses (see the “Methods” section for more details). 

\subsection{Laser-induced orbital magnetization without SOC.}



\begin{figure*}[ht!]
     \centering
      \includegraphics[width=1.0\textwidth]{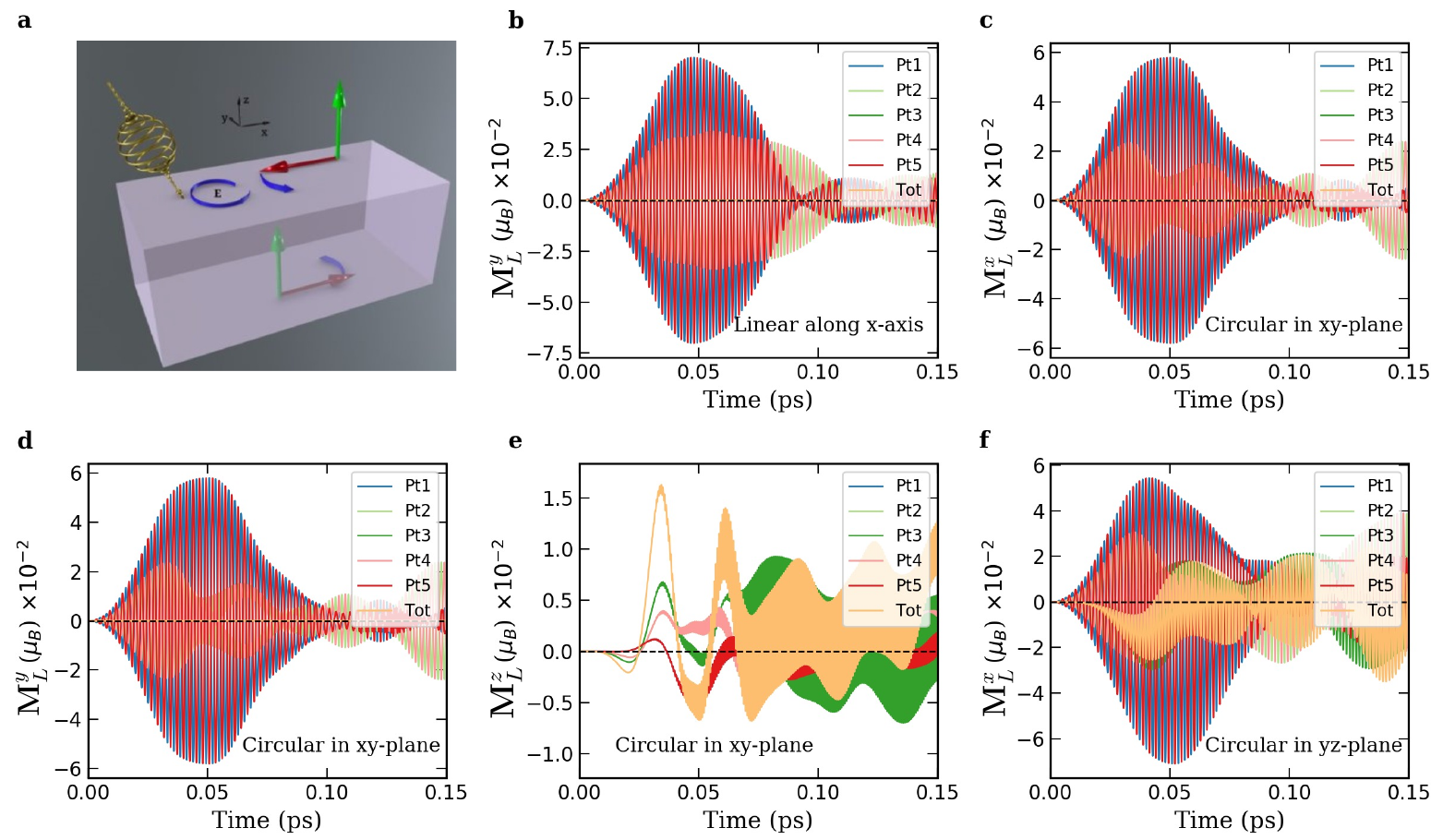}
      \caption{\textbf{Light-induced orbital magnetization in a Pt film in absence of spin-orbit coupling.} \textbf{a} Schematic representation of a laser pulse impinging on the surface of the Pt film. The electric field $\mathbf{E}$ impacts the electronic structure and can generate an orbital magnetization through orbital currents. These can accumulate via the orbital Hall effect (OHE) or through the inverse Faraday effect (IFE), illustrated by the red and green arrows respectively. 
      \textbf{b} $y$-component of the orbital moment induced by a laser pulse linearly polarised along the $x$-axis. This is the only non-vanishing component for this configuration.
      \textbf{c} $x$- \textbf{d} $y$- and \textbf{e} $z$-components of the orbital moment induced by a pulse with circular polarisation in the $xy$-plane. 
      \textbf{b-d} show components induced by the OHE while \textbf{f} shows the component expected from the IFE. \textbf{f} The $x$-component of the orbital moment induced by a circular pulse polarised in the $yz$-plane. This is the only non-vanishing component for this configuration, and includes contributions from both OHE and IFE.
      All pulses have an intensity of ~$\SI{9.7e8}{\volt\meter\tothe{-1}}$ and a width of $\SI{100}{\femto\second}$, with a laser frequency of the pulses have a fixed frequency $\omega=\SI{1.55}{\electronvolt}\hbar^{-1}$.}
     \label{fig1}
\end{figure*}


In Fig.~\ref{fig1}, we show the effect of light polarization on the induced orbital angular momentum for a (001)-oriented fcc Pt thin film consisting of 5 layers stacked along the $z$-direction in absence of SOC. The electric field couples to the linear momentum (and, therefore, the charge current), which affects the orbital momentum. As SOC is not present, the orbital and spin degrees are independent, and the light can only access the orbital angular momentum degree of freedom. For a pulse with linear polarization along the x-axis (i.e., the electric field $\mathbf{E}\parallel \mathbf{x}$), the charge current also oscillates along the x-axis and generates an orbital flow along the finite $z$ direction with $y$ polarization via the OHE. As the $z$ direction is finite, this orbital current accumulates on the two surfaces, as plotted as function of time in Fig.~\ref{fig1}b. There is also a flow of orbital current along the $y$ direction with $z$ polarization. However, as the system is infinite and periodic in the flow direction, it causes no measurable effect. Owing to the symmetry of the Pt film, the orbital angular momenta accumulating on opposite sides of the material are equal in magnitude and opposite in sign, resulting in zero net orbital angular momentum accumulation. This is expected from the OHE in a system with inversion symmetry. \new{We note that the laser-driven inverse Faraday effect is not operative for linear polarization of the laser.}

In Fig.~\ref{fig1}(c-e), the applied pulse has circular polarisation in the $xy$-plane, giving rise to oscillatory currents flowing in the same plane. Due to the OHE, an orbital current streaming along the z-axis has a polarization that oscillates in the $xy$-plane and accumulates on the edges, as displayed in Fig.~\ref{fig1}(c,d). As before, the total accumulation in the film sums up to zero due to the inversion symmetry. However, this case presents also a finite orbital moment along the $z$-axis that is induced by the IFE, as shown in Fig.~\ref{fig1}e. This component can reach up to about 0.015 $\MUB$ then fluctuates with time. \new{ In an intuitive picture the circularly-polarized laser induces a rotational motion with a fixed rotational sense during the pulse, and so the induced motion of the electronic charges is allowed to accumulate and lead to a non-zero average of the orbital magnetic moment. }

In Fig.~\ref{fig1}f, the plane of polarization is changed to $yz$. In this case, both OHE and IFE result in accumulations of the x component only: the first one originated on the flow of orbital current along the $z$ direction, and the second due to the circular polarization in the $yz$ plane. The charge flow along the $z$ direction does not generate any accumulation due to OHE, as the orbital currents flow in the $x$ and $y$ periodic directions. Since the signals are mixed, the total response does not cancel out as before. By comparing the maximum amplitude reached by the orbital moment $\mathrm{M}_L^z$ in Fig.~\ref{fig1}e to that reached by $\mathrm{M}_L^x$ in Fig.~\ref{fig1}f, we can in principle distinguish and potentially quantify the distinct impact of IFE and OHE.
\subsection{Laser-induced spin and orbital magnetization with SOC.}
\begin{figure*}[ht!]
     \centering
      \includegraphics[width=1.0\textwidth]{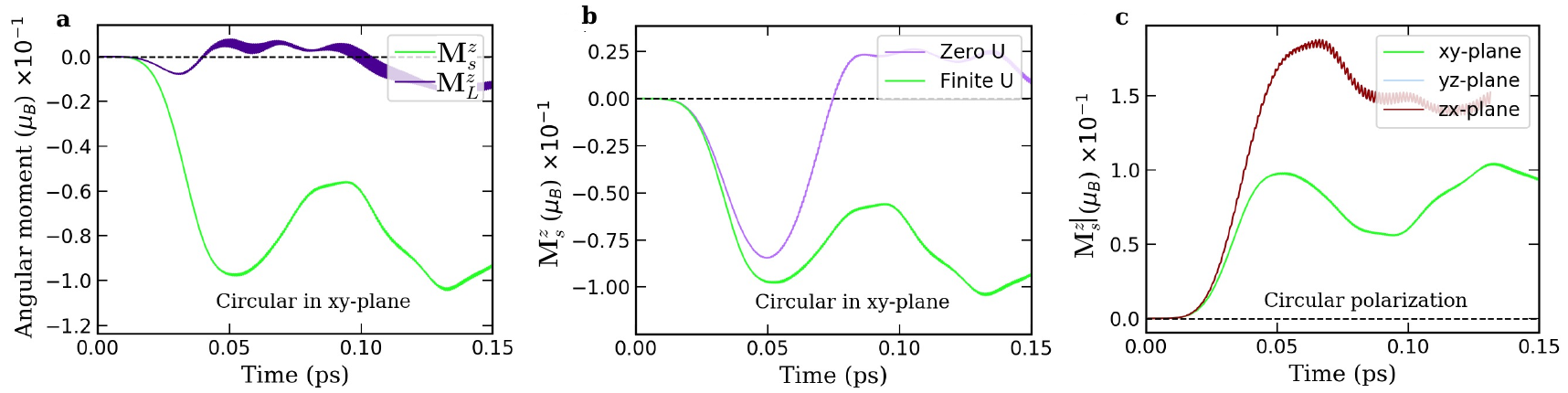}
      \caption{\textbf{Light-induced spin and orbital magnetizations in a Pt film incorporating spin-orbit coupling.}
      \textbf{a} The $z$-component of the total spin and orbital moment in the Pt film, which is  induced by the IFE when the laser pulse is polarized in the $xy$-plane. \new{\textbf{b}} The spin moment is also shown when  the effective electron-electron interaction strength $U=0$ (regular simulations assume $U=\SI{0.5}{\electronvolt}$ for Pt atoms). \new{\textbf{c}} A comparison between the absolute values of the Pt film spin moment as induced by a pulse polarized in the $xy$-, $yz$- and $zx$-planes. The $zx$-curve and $yz$-curve are identical, as expected from the symmetry of the system. All pulses have an intensity of ~$\SI{9.7e8}{\volt\meter\tothe{-1}}$.}
     \label{fig2} 
\end{figure*}

When SOC is present, a spin moment builds up following the light-induced orbital angular momentum.
As before, for a laser polarised in the $xy$-plane, the Hall effect (now spin and orbital) and IFE components are separated: while the former one induces circular oscillations of the spin and orbital moment vectors in the xy-plane with opposite directions in each surface of the system---therefore with vanishing total contribution---, the latter induces a finite net magnetic moment along the z direction.
Fig.~\ref{fig2}a  displays the total orbital $\mathrm{M}_L^z$ and spin $\mathrm{M}_s^z$ moments. Notice that even though the spin moment is indirectly caused by the orbital one via the spin-orbit coupling, the former reaches much higher values than the latter. 
The large value of the spin moment is also relatively stable, continuing close to $0.1\MUB$ up to $\SI{0.15}{\pico\second}$. This occurs due to the presence of the electron-electron interaction, taken into account effectively via the intra-atomic exchange interaction $U$, responsible for magnetism as expected from the \new{Stoner model~\cite{stoner1938collective}}. When this term is not present (i.e., $U=0$), the spin moment changes sign and reaches much smaller values at $\SI{0.15}{\pico\second}$ as shown in Fig.~\ref{fig2}\new{b}. 
By changing the polarization of the pulse to the $yz$ and $xz$-planes as shown in Fig.~\ref{fig2}\new{c}, one notices a large change, which can reach about 70\% enhancement, in the induced total spin moment.  This shows that a significant magnetic moment can be induced via a laser pulse in a non-magnetic material.  
\subsection{Impact of laser intensity on induced magnetizations.}

\begin{figure*}[ht!]
     \centering
      \includegraphics[width=1.0\textwidth]{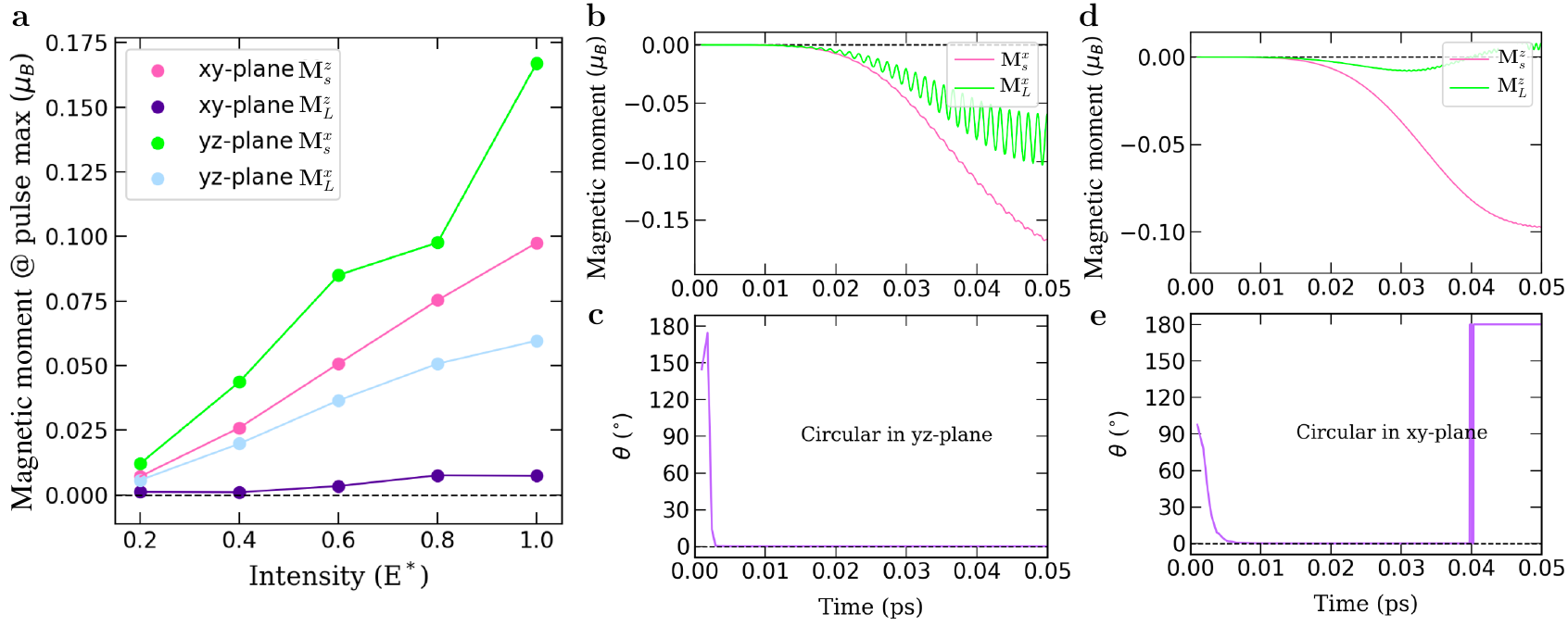}
      \caption{\textbf{Induced magnetizations at pulse peak.} \textbf{a} Comparison of the induced spin and orbital film magnetizations as  function of pulse intensity obtained at the pulse maximum for different plane polarizations ($xy$ and $yz$). 
      \textbf{b-c} example of the evolution of the spin and orbital magnetization for different polarizations of the pulses followed by the \textbf{c-d} angle between spin and orbital moment vectors. The reference value for the laser field intensity E$^*$ is given by ~$\SI{9.7e8}{\volt\meter\tothe{-1}}$.
           \label{fig3} }
\end{figure*}

To better understand how the discussed effects are generated by the electric field, we now look at how the pulse intensity affects the induced orbital and spin moments. Figure~\ref{fig3}a shows the values of the total norm of the moments  $\mathrm{M}_s$ and $\mathrm{M}_L$ at the pulse maximum ($\SI{50}{\femto\second}$)  as a function of the pulse intensity for circular polarizations in the $xy$- and $yz$-planes.  

At the peak of a pulse polarized in the $xy$-plane, the $z$-component of the induced orbital moment ($\mathrm{M}_L^z$) which is due to the IFE only is non-linear with the pulse intensity and saturates at 0.8 $\mathrm{E}^*$ (with $\mathrm{E}^* = \SI{9.7e8}{\volt\meter\tothe{-1}}$) while the $z$-component of the induced spin moment $\mathrm{M}_s^z$ follows a linear behavior. Note that at peak of the pulse, the spin magnetic moment does not reach its maximum (see e.g. Fig.~\ref{fig3}b), while the orbital moment can oscillate (Fig.~\ref{fig3}d). The $yz$-plane pulse induces moments in the $x$ direction. which are triggered by both the IFE and OHE. In this case, the orbital moment is relatively large, when compared to what is found with a pulse polarized in the  $xy$-plane.

By taking a close look at $\mathbf{M}_s$ and $\mathbf{M}_L$ in Figure~\ref{fig3}b-e induced by two different types of pulses, polarized in the $yz$ and $xy$-planes, we  notice that both magnetizations arise at about the same time. The orbital moment generated with the $xy$-plane polarized pulse is significantly smaller than that induced by the pulse polarized in the $yz$-plane. Looking at the angle between the two magnetizations, orbital and spin, one notices clearly that at short times after application of the laser, they are anti-parallel for the pulse polarized in the $yz$-plane while the angle goes down to 90 degrees for the $xy$-plane polarized pulse. This behavior does not last very long and the two magnetic moments quickly become parallel to each other, which could be induced by SOC. Note, however, that after 40 fs, the orbital magnetization induced by the $xy$-plane circularly polarized pulse experiences an oscillatory behavior which leads to an antipallel alignment with respect to the spin-magnetization.


\section*{Discussion}

In conclusion, we observe ultra-fast orbital and spin Hall effects in Pt layers. In ultra-fast experiments involving Pt, it is commonly employed as a substrate to facilitate angular momentum transfer due to its robust spin-orbit coupling (SOC). The sensitivity of these effects to light helicity and polarization offers insights into how the incorporation of such substrates may influence magnetization reversal at magnetic interfaces. Manipulating the polarization and helicity of laser pulses provides a potential avenue for distinguishing contributions to magnetization from the orbital and spin Hall effects against the backdrop of inverse Faraday contributions. Crucially, our predictions highlight a diverse range of mechanisms capable of inducing significant magnetic moments in initially non-magnetic materials. \new{In the future, it is appealing to explore the presence of phonons in affecting the emergence of the different unveiled ultrafast-induced magnetizations.}

We envision the potential use of transition metals, inherently non-magnetic, as a platform to explore unconventional magnetism at ultrafast timescales. This exploration holds promise for the implementation of all-optical addressed storage and memory devices.

\begin{methods}

\subsection{Theory.}\label{sec4.1}

We utilize a multi-orbital tight-binding Hamiltonian that takes into account the electron-electron interaction through a Hubbard like term and the spin-orbit interaction, as implemented on the TITAN code to investigate dynamics of transport and angular momentum properties in nanostructures~\cite{guimaraes2015fmafm,guimaraes2017dynamical,guimaraes2019comparative,guimaraes2020spin}. 
To describe the interaction of a laser pulse with the system, we include a time-dependent electric field described by a vector potential $\mathbf{A}(t) = -\int \mathbf{E}(t)dt $. The full Hamiltonian is given by
\begin{equation}
\mathcal{H}(t)= \mathcal{H}_\mathrm{kin} + \mathcal{H}_\mathrm{xc} + \mathcal{H}_\mathrm{soc} - \int\mathrm{d}\mathbf{r}\; \hat{\mathbf{J}}^\text{C}(\mathbf{r},t)\cdot \mathbf{A}(\mathbf{r},t) \;.
\end{equation}
More details on each term can be found in Ref.~\cite{Hamamera2022}.
The dipole approximation was used in the implementation of the vector potential, meaning that the spatial dependency is not included since the wavelength of the used light ($\hbar\omega = \SI{1.55}{\electronvolt} \rightarrow \lambda = \SI{800}{\nano\meter}$) is much larger than the lattice constant, and that the quadratic term as well as the other higher terms are zero~\cite{chen2019revealing}.

\subsection{Pulse shape.} \label{sec4.2}
For the right-handed circular pulse ($\sigma^+$) polarised in the $yz$-plane, for example, the pulse shape is described  using a vector potential of the following form~\cite{volkov2019attosecond},
\begin{equation}\label{eq:A(t)_circular}
\mathbf{A}(t) = -\frac{E_0}{\omega} \cos^2({\pi t}/{\tau}) [\sin(\omega t) \mathbf{\hat{y}} - \cos(\omega t) \mathbf{\hat{z}}],
\end{equation}
where $E_0$ is the electric field intensity, $\tau$ is the pulse width, $\omega$ is the laser central frequency which is set to $\SI{1.55}{\electronvolt}\hbar^{-1}$.
The magnetic field of the laser is neglected since it is much smaller than the electric field. The pulse width is assumed to be $\SI{100}{fs}$ for all calculations.

For the linear pulse ($\pi$) of a propagation direction along the diretion $\mathbf{\hat{u}}$, using the same central frequency, the vector potential is described as
\begin{equation}\label{eq:A(t)_linear}
\mathbf{A}(t) = -\frac{E_0}{\omega} \cos^2({\pi t}/{\tau}) \sin(\omega t) \mathbf{\hat{u}} \,.
\end{equation}

\subsection{Computational details.}
Calculations were performed on five layers of face-centered cubic platinum (Pt) stacked along the [001] direction using one atom per layer with the theoretical lattice constant of \SI{3.933}{\angstrom} given in Ref.~\citenum{Papaconstantopoulos:2015hr}, a uniform k-point grid of ($20\times 20$) and a temperature of \SI{496}{\kelvin} in the Fermi-Dirac distribution.
The initial step size for the time propagation is $\Delta t = 1$ a.u.\ which changes in the subsequent steps to a new predicted value such that a relative and an absolute error in the calculated wave functions stay smaller than $10^{-3}$~ \cite{Hairer2010}.

We tested the results for accuracy by increasing the number of k-points and decreasing the tolerance for the relative and absolute errors.
The method was also tested for stability by changing one of the laser parameters by a very small number while keeping the other parameters fixed, for one case that we already have results for.
The results then were not very different~\cite{higham2002accuracy}.

\end{methods}

\noindent \textbf{Code availability.} The tight-binding code that supports the findings of this study, TITAN, is  \new{ freely available on Ref.~\cite{titan_zenodo} or} from the corresponding author on request.

\noindent \textbf{Data availability.} The data that support the findings of this study are available from the corresponding authors on request.

\begin{addendum}

\item This work was supported by the Palestinian-German Science Bridge BMBF program and the European Research Council (ERC) under the European Union's Horizon 2020 research and innovation programme (ERC-consolidator Grant No. 681405-DYNASORE).
The authors gratefully acknowledge the computing time granted through JARA on the supercomputer JURECA\cite{jureca} at Forschungszentrum Jülich.

\item[Author contributions] 

\item[Competing Interests] The authors declare no competing interests.

\end{addendum}

\section*{References}
\bibliographystyle{naturemag}
\bibliography{references.bib}

\end{document}